\documentclass[aps,prb,twocolumn,showpacs]{revtex4}

\usepackage{epsfig}
\usepackage{graphics}
\usepackage{subfigure}
\usepackage{amssymb,amsmath}

\begin{document}

% \markboth{}{CONFIDENTIAL}

%\def\norma#1{\vert \vert {\bf #1} \vert \vert _2}
%\def\cond#1{cond({\bf #1})}
%\def\tcond#1{\widetilde {cond}({\bf #1})}
%\def\dprime#1{ #1^{\prime\prime}}
%\def\abs#1{\vert #1 \vert}
%\def\rerror#1{\norma{\delta #1}}
%\def\ferror#1{\frac{\norma{\delta #1}}{\norma{#1}}}
%\def\derror#1{{\norma{\delta #1}}/{\norma{#1}}}
%\def\TR#1{\parallel \!#1\!\parallel^2}

\def\e{\begin{equation}}
\def\f{\end{equation}}
\def\_#1{{\bf #1}}
\def\o{\omega}
\def\.{\cdot}
\def\x{\times}
\def\E{\epsilon}
\def\H{\Upsilon}
\def\A{\alpha}
\def\B{\beta}
\def\O{\Xi}
\def\M{\mu}
\def\D{\nabla}
\newcommand{\ds}{\displaystyle}
\def\l#1{\label{eq:#1}}
\def\r#1{(\ref{eq:#1})}
\def\=#1{\overline{\overline #1}}
\def\##1{{\bf#1\mit}}

\title{Lateral-drag Casimir forces induced by anisotropy}

 \author{Igor S. Nefedov$^{1,2}$, J. Miguel Rubi$^{3}$}
 \affiliation{$^1$Aalto University,
 School of Electrical Engineering,
P.O. Box 13000, 00076 Aalto, Finland\\
 $^2$Laboratory Nanooptomechanics, ITMO University, St. Petersburg, 197101, Russia
\\$^3$Statistical and Interdisciplinary Physics Section, Departament de Fisica de la Mat\`{e}ria Condensada, Universitat de Barcelona, Marti i Franqu\`{e}s 1, 08028 - Barcelona, Spain}

 \date{\today}

 \begin{abstract}
We predict the existence of lateral drag forces near the flat surface of an absorbing slab of an anisotropic material. The forces originate from the fluctuations of the electromagnetic field, when the anisotropy axis of the material forms a certain angle with the surface. In this situation, the spatial spectra of the fluctuating electromagnetic fields becomes asymmetric, different for positive and negative transverse wave vectors components.  Differently from the case of van der Waals interactions in which the forward-backward symmetry is broken due to the particle movement or in quantum noncontact friction where it is caused by the mutual motion of the bodies, in our case the lateral motion results merely from the anisotropy of the slab. This new effect, of particular significance in hyperbolic materials, could be used for the manipulation of nanoparticles.
 \end{abstract}

 \maketitle

% \section{Introduction}

%%%%%  New
Fluctuating electromagnetic fields are responsible for important phenomena such as thermal emission, radiative heat transfer, van der Waals interactions,  Casimir effect, and van der Waals friction between bodies \cite{volokitin}.
The existence of attractive forces
between two perfectly conducting parallel plates,
induced by vacuum fluctuations at zero temperature,
was predicted by Casimir in 1948 \cite{Casimir, Lamoreaux} and subsequently by Lifshitz \cite{Lif} for any media at finite temperature.
A general electromagnetic fluctuation theory, referred to as fluctuational electrodynamics, was proposed by Rytov in 1950s \cite{Rytov}.
The conventional Casimir force between two parallel surfaces bounding a vacuum gap is  orthogonal to the surfaces, of attractive nature when the separating distance is small and the interaction is due to inhomogeneous fields of evanescent waves, excited by fluctuating currents. For larger vacuum gap widths, the Casimir force becomes repulsive \cite{Antezza,Rubi}.

The lateral component of the Poynting vector, integrated over the whole spatial spectrum, vanishes near flat surfaces because its positive and negative components balance each other out. This symmetry can be broken by a mutual lateral movement of the bodies, as happens in the  case of contact-free van der Waals and quantum friction \cite{Pendryfric,Volfric}.
To observe these forces, one applies an electric current in a conducting layer and measures  the friction drag effect of electrons in a second parallel metallic layer \cite{Pogreb,Price}.
Lateral drag forces can also be observed nearby surface inhomogeneities, such as corrugations.
These forces, however, cause local displacements related to the periodicity of the corrugations \cite{Chen,Bimonte,Rodriguez} and not a net movement over an appreciable distance. A lateral propulsion force, exerted on an anisotropic particle, was predicted by M\"{u}ller and Kr\"{u}ger \cite{Muller}.

%%%%  END New

In this work, we propose a new mechanism able to generate lateral forces.
If the absorbing medium is anisotropic and the anisotropy axis is tilted with respect to the slab surface, absorption of the TM-polarized wave incident on the slab is different for  positive and negative incident angles, although the reflection be the same \cite{Has,ScRep}.

\begin{figure}[h!]
\centering \epsfig{file=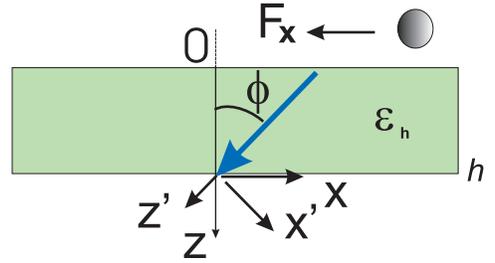, width=0.35\textwidth}
\caption{The anisotropic slab and the small particle affected by the lateral force. The anisotropy axis is indicated by a thick blue arrow. The black arrow shows the direction of motion of the particle due to the action of the $x$-component of the radiative forces $\_F_x$.}
\label{slab}\end{figure}

We solved the boundary-value problem for electromagnetic waves, excited by point-like fluctuating currents within a finite-thickness slab of an anisotropic medium (see Fig.~\ref{slab}).
Due to the homogeneity of the considered geometry in the $x$ and $y$ directions, the electric and magnetic fields and the current can be represented by means of their corresponding Fourier transforms $\_E(\o,k_x,k_y)$, $\_H(\o,k_x,k_y)$, and $\_j(\o,k_x,k_y)$.
%\e\l{F}\left[\begin{array}{l}
%\_E(\o,x,y)\\
%\_H(\o,x,y)\\
%\_j(\o,x,y)\end{array}\right]
%=\int\int_{-\infty}^{\infty}
%\left[\begin{array}{l}
%\_E(\o,k_x,k_y)\\
%\_H(\o,k_x,k_y)\\
%\_j(\o,l_x,k_y)\end{array}\right]
%\frac{e^{i(k_xx+k_yy)}}{(2\pi)^2}\,dk_x\,dk_y,
%\f
To find a fully accurate solution of the electromagnetic fields is a difficult task because the fields in the considered geometry cannot be decomposed into TM and TE waves.
To show the existence of a lateral force, however, it is enough to consider TM waves propagating along the slab in the anisotropy plane, assuming $k_y=0$.

% \section{Waves in anisotropic media}

In the coordinate system $(x',y',z')$ (see Fig.~\ref{slab}) the relative permittivity tensor has the diagonal form
\e\l{p1} \=\E'=\E_{\parallel}\_z'_0\_z'_0+\E_t(\_x'_0\_x'_0+\_y'_0\_y'_0).\f
where the subscript $0$ denotes unit vector. The components of the permittivity tensor in the reference frame associated with the slab interface
are given in the Supplemental Material.
If the anisotropy axis is tilted with respect to the slab interfaces, the Maxwell equations can be split up into TM and TE subsystems, provided that the wave vector lies in the anisotropy axis plane or is orthogonal to it. We will restrict our analysis to TW waves.

The propagation constants of those waves, traveling along the $z$-direction for a fixed $k_x$ are given by \cite{Has}
\e\l{e8}
k_z^{(1,2)}=\frac{-k_x\E_{xz}\pm\sqrt{(\E_{xz}^2-\E_{xx}\E_{zz})(k_x^2-k_0^2\E_{zz})}}{\E_{zz}}.\f
where $k_0$ is the wavenumber in vacuum.

The transverse wave impedance $Z_{1,2}$, connecting tangential field components, reads \cite{Has}
\e\l{a2}
Z_{1,2}=\frac{E_x}{H_y}=\pm
\frac{\eta}{k_0}\frac{\sqrt{k_x^2-k_0^2\E_{zz}}}{\sqrt{\E_{xz}^2-\E_{xx}\E_{zz}}},
\f
where $\eta=120\pi$\,Ohm is the wave impedance of vacuum.
%The sign in \r{a2} is chosen by such a way that ${\rm Re}(Z_{1,2})>0$ for waves propagating in the %positive $z$-direction.

%\section{Fields produces by fluctuating currents}

For the tangential field components ${\rm X}(z)=\left(E_x(z),H_y(z)\right)$, excited by the fluctuating currents $j_x(z),\,j_z(z)$ located within the absorptive layer $0<z<h$ (see Fig.~\ref{slab}),
the Maxwell equations reduce to the system of two ordinary differential equations:
\e\l{me}
\frac{d}{dz}{\rm X}(z)=[{\rm A}] {\rm X}(z)+{\rm F}(z)\f
where the matrix elements of [A] are given by
\e\l{amat}\begin{array}{lr}
A_{11}=-i\frac{k_x\E_{xz}}{\E_{zz}},  & A_{12}=i\eta k_0\left(1-\frac{k_x^2}{k_0^2\E_{zz}}\right)\\
A_{21}=i\frac{k_0}{\eta}\left(\E_{xx}-\frac{\E_{xz}^2}{\E_{zz}}\right), &
A_{22}=-ik_x\frac{\E_{xz}}{\E_{zz}}
\end{array}\f
and the components of the vector F($z$)=$(F_1(z),F_2(z))$ are
\e\l{cur}
\begin{array}{l}
F_1(z)=\eta\frac{k_x}{k_0\E_{zz}}j_z(z)=aj_z(z), \\
F_2(z)=\frac{\E_{xz}}{\E_{zz}}j_z(z)-j_x(z)=bj_z(z)-j_x(z).
\end{array}\f
The elementary bulk current source has the form: $\_j(z)=\_j_0(z')\delta(z-z')$.

The solution of Eq.~\r{me} for points $0<z<h$ is \cite{Lankaster}:
\e\l{c}
{\rm X}(z)=e^{[{\rm A}]z}{\rm X}(0)+\int_0^z{e^{[{\rm A}](z-\tau)}{\rm F}(\tau)}d\tau,\f
with $[{\rm M}(z)]=e^{[{\rm A}]z}$  the transfer matrix. Expressions for the $2\times 2$ matrix components for the case in which the wave impedances and vector components are different for waves propagating in opposite directions are given in \cite{Has,ScRep}.
Taking into account that $Z_2=-Z_1=Z$, those expressions reduce to
\e\l{tm}\begin{array}{lr}
M_{11}(z)=\frac{1}{2}\left[e^{ik_{z1}z}+e^{ik_{z2}z}\right] &
M_{12}(z)=\frac{Z}{2}\left[e^{ik_{z1}z}-e^{ik_{z2}z}\right] \\
M_{21}(z)=\frac{1}{Z}\left[e^{ik_{z1}z}-e^{ik_{z2}z}\right] &
M_{22}(z)=M_{11}(z).\end{array}\f

The boundary conditions are:
$X_2(0)=X_1(0)/Z_0,\;X_2(h)=-X_1/Z_0$, where $Z_0=\eta\sqrt{(k_0^2-k_x^2)}/k_0$ is the {\it transverse} wave impedance in vacuum.
We can then express the tangential field components at the interface $x=0$, created by a current located at $z'$ in the form
\e\l{x1}\begin{array}{l}
X_1(0,z')=\frac{1}{\Delta}\int_0^h\left[U(\tau)j_{x0}(z')+\right.\\
\left.V(\tau)j_{z0}(z')\right]\delta(\tau-z')\,d\tau=\\
=\frac{1}{\Delta}\left[U(z')j_{x0}(z')+V(z')j_{z0}(z')\right],\end{array}
\f
where
\e\l{delta}\begin{array}{l}
\Delta=M_{11}(h)+M_{22}(h)-M_{12}(h)/Z_0-M_{21}(h)Z_0, \\
U(\tau)=M_{12}(h-\tau)-Z_0M_{22}(h-\tau),\\
V(\tau)=a[Z_0M_{21}(h-\tau)-M_{11}(h-\tau)]+\\
+b[Z_0M_{22}(h-\tau)-M_{12}(h-\tau)].
\end{array}
\f

The Fourier components of the electric and magnetic fields out of the layer are given by
\e\l{EH}
E_x(k_x,z,z')=\left\{\begin{array}{l} X_1(0,z')e^{-ik_{z0}z},\; z<0, \\
X_1(h,z')e^{i k_{z0}(z-h)},\; z>h,\end{array}\right.
\f
where $k_{z0}=\sqrt{k_0^2-k_x^2}$ and $H_y(k_x,z,z')=E_x(k_x,z,z')/Z_0$.
For evanescent waves, $|k_x|>k_0$, we have to take $k_{z0}=-i|k_{z0}|$, if $z<0$
and $k_{z0}=i|k_{z0}|$, if $z>h$.

The average values of the fluctuating currents vanish, only their correlations contribute to the energy flux. These correlations are given through the fluctuation-dissipation theorem \cite{Landau}
\e\l{fdt}
\begin{array}{lc}
\langle j_m(\_r,\omega)j^*_n(\_r',\omega')\rangle=  \\
 \frac{4}{\pi}\omega\E_0\E_{mn}''(\omega)\delta(\_r-\_r')\delta(\omega-\omega')\Theta(\omega,T),
\end{array}\f
with $\vec r=(x,z)$ and
\e\l{Th}
\Theta(\o,T)=\frac{\hbar\o}{2}+\frac{\hbar\o}{e^{\hbar\o/(k_BT)}-1}
\f
the Planck's oscillator energy.
In Eq.~\r{fdt}  $\E_{mn}''\equiv {\rm Im}(\E_{mn})$, $\E_0$ is the permittivity of vacuum, $\hbar$ the reduced Planck constant,
$T$ the temperature, and $k_B$ the Boltzmann constant.

% \section{The lateral Poynting vector}

The ensemble-averaged Poynting vector in the plane $z=0$, for the $k_{x}$ mode,
induced by fluctuating currents located within the slab,  $0<z',z''<h$, reads
\e\l{P}
\langle S_z(\o,k_x)\rangle=\frac{1}{2}\int_0^h\int_0^h\langle E_x(k_x,z')H_y^*(k_x,z'')\rangle\,dz'dz''
 \f
%Here we also used the ergodic hypothesis \cite{Vol1}:
%\e\l{erg}
%\langle\_S^1_z(\_r,\omega)\rangle= \int^{\infty}_0\frac12\langle
%E_x(\_r,\omega) H_y^*(\_r,\omega')\rangle d\omega'
% \f
Using the fluctuation-dissipation theorem \r{fdt}, we then obtain
\e\l{Sz}
\langle S_z(\o,k_x)\rangle=
\frac{4\o\E_0\Theta(\o,T)}{2\pi|\Delta|^2Z_0^*}\left[D_1\E_{xx}''+D_2\E_{zz}''+
D_3\E_{xz}''\right],
\f
where the coefficients $D_1$, $D_2$ and $D_3$ are given in the Supplemental Material.
%\e\l{D1}\begin{array}{l}
%D_1=\int_0^hU(\tau)U^*(\tau)\,d\tau,\\
%D_2=\int_0^hV(\tau)V^*(\tau)\,d\tau,\\
%D_3=\int_0^hU(\tau)V^*(\tau)\,d\tau+c.c,
%\end{array}\f

If the anisotropy axis is parallel or orthogonal to the interface, then $\langle S_z(\o,k_x)\rangle=\langle S_z(\o,-k_x)\rangle$.
Otherwise, that average becomes asymmetric with respect to the normal to the slab interface, as occurs in absorption \cite{Has,ScRep}.
Outside the slab the lateral time-averaged component of the Poynting vector is given by
\e\l{Sx}
\langle S_x(\o,k_x,z)\rangle=-\frac{1}{2}E_zH_y^*=\frac{k_x}{k_{z0}}\langle S_z(\o,k_x)\rangle f(z),
\f
where $f(z)=1$, if $|k_x|<k_0$, and $f(z)=e^{2|k_{z0}|z}$ $(z<0)$, if $|k_x|>k_0$.

The contribution of all $k_x$-modes to the $x$-component of the Poynting vector is
\e\l{all}
\langle S_x(\omega)\rangle=\frac{1}{2\pi}\int_{-\infty}^{\infty}\,\langle S_x(\o,k_x,z)\rangle\,dk_x.\f
Since $\langle S_x(\o,k_x,z)\rangle\neq \langle S_x(\o,-k_x,z)\rangle$, one can expect the appearance of radiative forces  dragging a particle placed nearby the anisotropic slab along its surface.

%Fig.~\ref{Px} illustrates spatial spectral dependence of the contribution
%$\Delta S_x(k_x)=\langle S_x(\o,k_x,z)\rangle+\langle S_x(\o,-k_x,z)$.
%\begin{figure}[h!]
%\centering \epsfig{file=Sx.eps, width=0.5\textwidth}
%\caption{$\Delta S_x(k_x)$ versus $k_x$. Components of the permittivity tensor in the proper coordinate %system, where tensor is diagonal, are the following: $\E_{\parallel}=1+i0.2$,
%$\E_{\perp}=1$. Thickness of the slab $h=3\,\mu$m.
%Black, blue, and red curves correspond to the tilt angles $\phi=5^{\circ},\,
%10^{\circ}$, and $15^{\circ}$, respectively.
%}
%\label{Px}\end{figure}
%Evidently, integration of $\Delta S_x(k_x)$ over $k_x$ from 0 to $\infty$ will give nonzero result.

When $|k_x|<k_0$, expression \r{P} gives us the thermal power radiating from the slab in the optical axis $(x,z)$ plane, i.e. $k_y=0$.
The total energy flux density in the $x$-direction, produced by electromagnetic fluctuations,
is given by
\e\l{tot}
\langle S_x(z)\rangle^{\rm tot}=\int_0^{\infty}
\frac{d\o}{(2\pi)^3}\int\int_{-\infty}^{\infty}
\langle S_x(\o,k_x,k_y)\rangle f(z)\,dk_xdk_y
\f
where $k_{z0}=\sqrt{k_0^2-k_x^2-k_y^2}$. An exact value of this quantity for nonzero values of $k_x$ and $k_y$ is difficult to obtain since the fluctuating fields in the slab are carried by hybrid waves whose solution is more difficult to obtain than that for TM waves.

The calculation of the total energy flux in the $x$-direction is based on the following consideration. If the optical axis is orthogonal to the slab interfaces or if the medium is isotropic,
due to azimuthal symmetry, one can replace $dk_xdk_y$ by $2\pi k_xdk_x$, therefore
\e\l{id}
\int_{-\infty}^{\infty}\int_{-\infty}^{\infty}\langle S_{z}(\o,k_x,k_y)\rangle\,dk_x\,dk_y=
2\pi\int_0^{\infty}\langle S_{z}(\o,q)\rangle q\,dq
\f
and $\langle S_x(z)\rangle^{\rm tot}=\langle S_x^s(z)\rangle+\langle S_x^p(z)\rangle\equiv 0$.
Let us consider separately the cases in which the wave vector is either parallel or orthogonal to the plane normal to the slab surface, containing the anisotropy axis.
In both cases, waves in an anisotropic slab can be split up into $p$-polarized and $s$-polarized waves. Obviously,
for the $s$-polarized waves the asymmetry never takes place and $\langle S_{x}^{p}(z)\equiv 0$ for both cases, $k_x\neq 0,\,k_y=0$ and $k_x=0,\,k_y\neq 0$.
For $p$-polarized waves, the asymmetry is absent if $k_x=0$, at any $k_y$ and it is
maximal if $k_x\neq 0,\,k_y=0$. Thus, the
integral over waves, propagating in the $y$-direction, i.e. $k_x=0$,  gives zero contribution to
$\langle S_p(z)\rangle_x$.
A good approximation to the total lateral energy flux density is then given by
\e\l{fin}
\langle S_x^{\rm appr}(\o,z)\rangle^{\rm tot}\approx \frac{1}{4\pi}\int_0^{\infty}\,d\o\int_{0}^{\infty}
\langle S_x^p(\o,q)\rangle f(z)q\,dq.
\f

% \section{Absorption and the lateral Poynting vector in graphene-based hyperbolic structure}

Asymmetry with respect to $\pm k_x$ takes place for any absorbing anisotropic material but it becomes particularly important for media characterized by a hyperbolic dispersion, for the so-called {\itshape hyperbolic materials,}
whose diagonal components of the permittivity tensor have different signs.
To illustrare the lateral drag effect, we will consider the orthorhombic modification of boron nitride which exhibits hyperbolic dispersion in certain frequency ranges \cite{Ordin,Felinskyi}.
The components of the permittivity tensor are given by the Lorentz model:
\e\l{lor}
\E_{\parallel,\perp}=\E_{\parallel,\perp}^{\infty}+
\frac{S_{\parallel,\perp}(\o_{\parallel,\perp}^{\tau})^2}
{(\o_{\parallel,\perp}^{\tau})^2-\o^2-i\o\Gamma_{\parallel,\perp}},
\f
where $\o_{\parallel,\perp}^{\tau}$ and $S_{\parallel,\perp}$ are, respectively, the transverse phonon frequency and the oscillator strength of the lattice vibration for the parallel and perpendicular polarizations, and $\Gamma_{\parallel,\perp}$ is the damping constant. The constants $\E_{\parallel,\perp}^{\infty}$ are the components of the permittivity tensor at frequencies $\o$ that greatly exceed the phonon resonance frequency $\o_{\parallel,\perp}^{\tau}$.
The values of the parameters of \r{lor} used are: $\E_{\parallel}^{\infty}=2.7$,
$S_{\parallel}=0.48$, $\o_{\parallel}^{\tau}=1.435\times 10^{14}$\,rad/s,
$\Gamma_{\parallel}=8.175\times 10^{11}$\,rad/s,
$\E_{\perp}^{\infty}=5.2$,
$S_{\perp}=2$, $\o_{\perp}^{\tau}=2.588\times 10^{14}$\,rad/s,
$\Gamma_{\perp}=1.29\times 10^{12}$\,rad/s.
For this parameters, the Lorentzian resonances of $\E_{\parallel}$ and $\E_{\perp}$,
take place at frequencies $\approx 22.8$\,THz and $\approx 41.2$\,THz, respectively.

\begin{figure}[h!]
\centering \epsfig{file=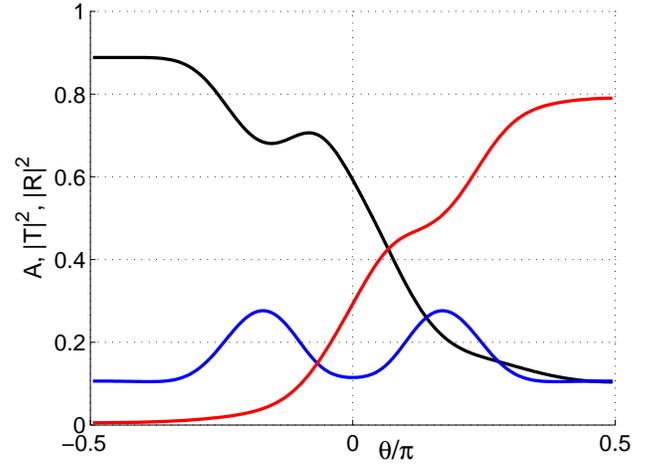, width=0.5\textwidth}
\caption{Absorption (black), transmission (red) and reflection (blue) of the plane wave incident onto the slab of orthorhombic boron nitride, versus the incidence angle.
The thickness of the layer is 1.5\, $\mu$m and the tilt angle $\phi=40^{\circ}$.
}
\label{angle}\end{figure}

The dependence of the transmission, $|T|^2$, $|R|^2$, and absorption $A=1-|T|^2-|R|^2$ on the incidence angle, calculated at 45\,THz which is close to the Lorentzian resonance for $\E_{\parallel}$,  is shown in Fig.~\ref{angle}.
The displayed dependence of the absorption versus the incidence angle is the signature of the asymmetry of thermal emission and radiative forces with respect to $k_x$.

% \section{Radiative forces on a dipole particle}

To evaluate the effect of the lateral forces, we will consider a small particle moving under the influence of fluctuating eelectromagnetic fields.
The dipolar force can be written as \cite{Silvia}
\e\l{for}\begin{array}{l}
\langle\_F\rangle=\frac{1}{4}{\rm Re}\{\alpha\}\nabla|\_E|^2+
\sigma\frac{1}{2}{\rm Re}\left\{\frac{1}{c}
\_E\times\_H^*\right\}+\\
+\sigma\frac{1}{2}{\rm Re}\left\{i\frac{\E_0}{k_0}(\_E\cdot\nabla)\_E^*\right\}
\end{array}\f
where $\alpha$ is the polarizability of the particle given by

\e\l{polar}\begin{array}{ll}
\alpha=\frac{\alpha_0}{1-i\alpha_0k_0^3/(6\pi\E_0)}, &
\alpha_0=4\pi\E_0r^3\frac{\epsilon-1}{\epsilon+2},
\end{array}\f
with $r$ and $\epsilon$ its radius and permittivity, respectively, and $\sigma=k_0{\rm Im}\{\alpha\}/\E_0$.

The first term in \r{for}, related to the gradient forces, causes attraction of the particle toward the slab interface due to the $z$-dependence of fields through $e^{|k_{z0}|z}$, for $z<0$ (van-der-Waals forces \cite{volokitin}).
The explicit expression for $\nabla |\_E|^2$ is
\e\l{grad}\begin{array}{l}
\nabla |\_E(\omega,k_x,z)|^2=\frac{\partial}{\partial z}f(z)\left[\langle E_xE_x^*\rangle+\langle E_zE_z^*\rangle\right]=\\
=2\langle S_z(\omega,k_x)\rangle\eta f(z)\frac{2k_x^2-k_0^2}{k_0},\end{array}\f
Only the evanescent waves $(|k_x|<k_0)$ contribute to this force.

In the second contribution, the $x$ and $z$-component of the Poynting vector exert pulling forces along the corresponding directions. At small $|z|$, the attractive gradient force is dominant, whereas at larger $|z|$ the dominant force is the repulsive force proportional to the $z$ component of the Poynting vector.

Due to the fact that

\e\l{fxz}\begin{array}{l}
\langle (\_E\cdot\nabla)\_E^*\rangle_x=E_x\frac{\partial}{\partial x}E_x^*+
E_z\frac{\partial}{\partial z}E_x^*=0,\\
\langle (\_E\cdot\nabla)\_E^*\rangle_z=E_x\frac{\partial}{\partial x}E_z^*+
E_z\frac{\partial}{\partial z}E_z^*=0,
\end{array}\f
the third term in \r{for} does not contribute to the radiative forces.
The $x$-component of the Poynting vector in \r{for} is given by Eq.~\r{fin} and the $z$-component can be found by replacing $\langle S_x^p(\o,q)\rangle$ by $\langle S_z^p(\o,q)\rangle=-(k_{z0}/k_x)\langle S_x^p(\o,q)\rangle$.

Note, that since the TM (p)-polarized waves only contribute to the lateral Poynting vector at $k_y=0$ and in our approximation we have not taken into account the hybrid nature of waves in the anisotropic slab at $k_y\neq 0$, we can consider contributions of the p-polarized waves only to the $z$-directed forces.

As an example of particle experiencing a lateral drag force, we consider a spherical gold nanoparticle whose complex permittivity $\E_g$ in the infrared and far infrared ranges is, according to the Drude model, given by
\e\l{Drude}
\E_g=\E_{\infty}-\frac{\omega_p^2}{\omega^2+i\omega\omega_r},
\f
where $\omega_p=1.367\times 10^{16}$\,rad/s, and $\omega_r=10.5\times 10^{13}$\,rad/s are the plasma frequency and the damping frequency, respectively, and $\E_{\infty}=9.5$ \cite{Johnson}. The radius of the particle is 15\,nm.

\begin{figure}[h!]
\centering \epsfig{file=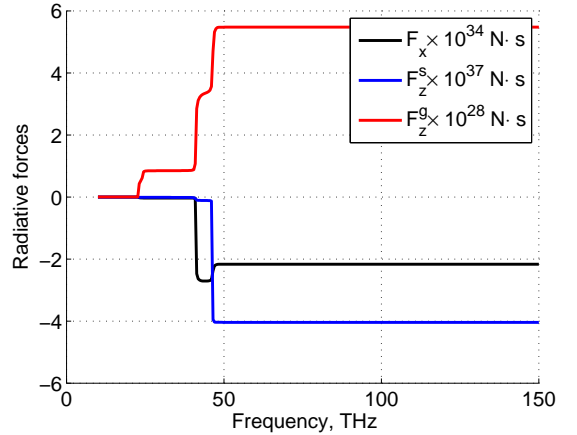, width=0.45\textwidth}
\caption{Radiative forces versus frequency for $h$=400\,nm, $T$=450\,K, and $\phi$= 50$^{\circ}$.}
\label{force}\end{figure}

Fig.~\ref{force} shows the  spectral densities of the dipolar forces acting on the particle,  computed at $z=30$\,nm.
The normal component of the force consists of contributions from the gradient force, $F^g_z(\o)$, and the force, caused by the Poynting vector, $F^s_z(\o)$.
One can verify, that at such a distance the van-der-Waals attractive force is dominant.
Nevertheless, the magnitude of the lateral drag force is even much stronger than the repulsive $z$-directed force. The reason is the fact that $F^s_z$ is caused by propagating waves only, while $F_x(\o)$ includes contributions from both propagating and evanescent waves.
The spectral densities of these forces were calculated within the frequency range 1\,THz--250\,THz in which the Lorentz model for orthorhombic boron nitride \r{lor} seems to be applicable, and pictured within 10\,THz--150\,THz in Fig.~\ref{force}.
The corresponding values of the forces, integrated over frequency, are :
$F^g_z=2.7\times 10^{-15}$\,N, $F^s_z=-4.2\times 10^{-25}$\,N, $F_x=-2.2\times 10^{-21}$\,N.
Our estimations thus show that the predicted lateral force could be detected experimentally.
Under this lateral force, the nanoparticle will move with acceleration of 0.81\,sm/s$^2$.
One can expect saturation of the particle speed due to contact-less quantum friction.

%\section{Conclusion}

In summary, we have predicted a new physical effect caused by  fluctuations of the electromagnetic field nearby an absorbing anisotropic slab: the presence of lateral drag forces emerging when the anisotropy axis of the slab is tilted. This effect systematically occurs in any absorbing anisotropic media, but it may be especially noticeable for materials with a strong anisotropy.
To prove the existence of such forces, we have solved the boundary value problem in the TM-waves approximation that ignores the hybrid nature of the waves supported by the slab for the considered anisotropy, if $k_y\neq 0$.
The presence of these drag forces which can be referred to as ``{\itshape the driving force from nothing}" \cite{Lambr} can play an important role in the manipulation of nanoparticles close to a surface.

\end{document}

% \markboth{}{CONFIDENTIAL}

%\def\norma#1{\vert \vert {\bf #1} \vert \vert _2}
%\def\cond#1{cond({\bf #1})}
%\def\tcond#1{\widetilde {cond}({\bf #1})}
%\def\dprime#1{ #1^{\prime\prime}}
%\def\abs#1{\vert #1 \vert}
%\def\rerror#1{\norma{\delta #1}}
%\def\ferror#1{\frac{\norma{\delta #1}}{\norma{#1}}}
%\def\derror#1{{\norma{\delta #1}}/{\norma{#1}}}
%\def\TR#1{\parallel \!#1\!\parallel^2}

\def\e{\begin{equation}}
\def\f{\end{equation}}
\def\_#1{{\bf #1}}
\def\o{\omega}
\def\.{\cdot}
\def\x{\times}
\def\E{\epsilon}
\def\H{\Upsilon}
\def\A{\alpha}
\def\B{\beta}
\def\O{\Xi}
\def\M{\mu}
\def\D{\nabla}
\newcommand{\ds}{\displaystyle}
\def\l#1{\label{eq:#1}}
\def\r#1{(\ref{eq:#1})}
\def\=#1{\overline{\overline #1}}
\def\##1{{\bf#1\mit}}

\title{Lateral-drag Casimir forces induced by anisotropy. Supplemental Material}

 \author{Igor S. Nefedov$^{1,2}$, J. Miguel Rubi${^3}$}
 \affiliation{$^1$Aalto University,
 School of Electrical Engineering,
P.O. Box 13000, 00076 Aalto, Finland\\
 $^2$Laboratory Nanooptomechanics, ITMO University, St. Petersburg, 197101, Russia
\\$^3$Statistical and Interdisciplinary Physics Section, Departament de Fisica de la Mat\`{e}ria Condensada, Universitat de Barcelona, Marti i Franqu\`{e}s 1, 08028 - Barcelona, Spain}


\maketitle

\section{Permittivity tensor}

In the coordinate system $x,y,z$, associated with the slab interface,
the components of the permittivity tensor $\=\E$ are the following \cite{Has}:
\e\l{e5}\begin{array}{l}
\E_{xz}=\E_{zx}=(\E_{\parallel}-\E_t)\cos{\phi}\sin{\phi}\\
\E_{xx}=\E_{\parallel}\sin^2{\phi}+\E_t\cos^2{\phi}\\
\E_{zz}=\E_{\parallel}\cos^2{\phi}+\E_t\sin^2{\phi}.\end{array}\f
If the anisotropy axis is tilted with respect to the slab interfaces, the Maxwell equations split up into TM and TE sub-systems provided that the wave vector lies in the anisotropy axis plane or is orthogonal to it. In the considered case of electrically anisotropic materials, the TM waves are the extraordinary ones.
\begin{figure}[h!]
\centering \epsfig{file=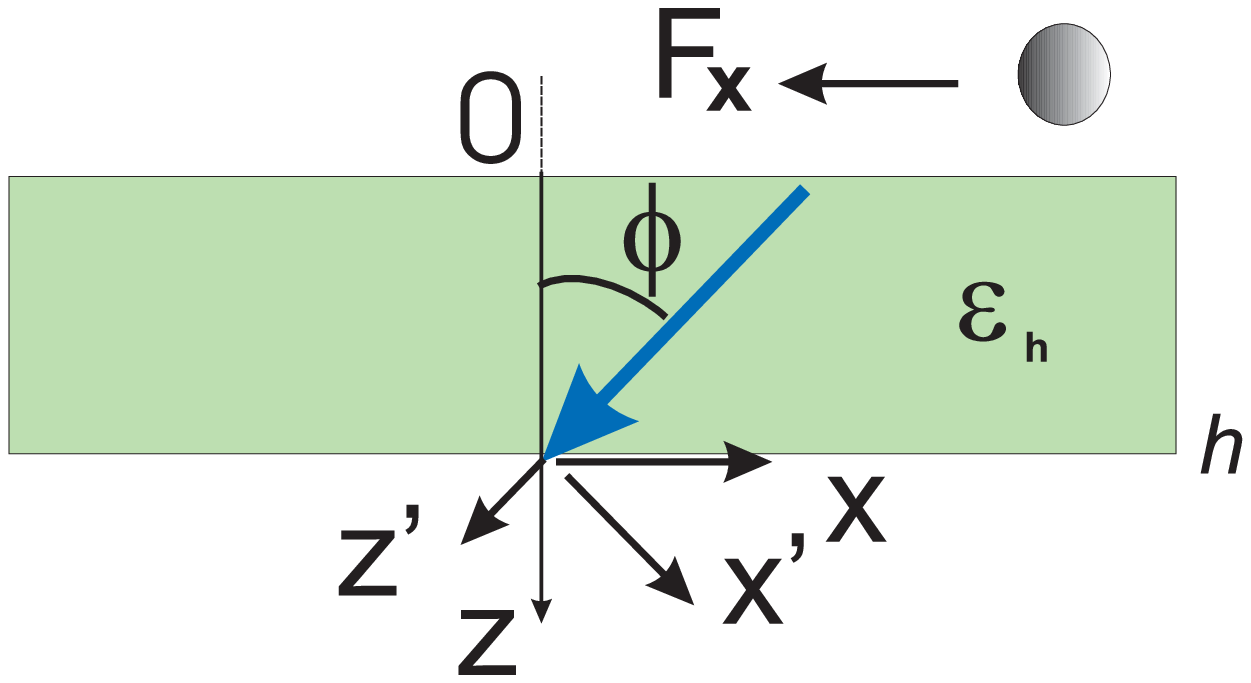, width=0.35\textwidth}
\caption{The anisotropic slab with a small particle close to it. The anisotropy axis is shown with a thick blue arrow. The arrow indicates the $x$-component of the radiative forces $\_F_x$.}
\label{slab}\end{figure}

\section{Eigenwaves in the boron nitride medium}

The propagation constants and the transverse wave impedance of the TM waves, traveling along the $z$-direction under fixed $k_x$, are given by Eq.~(2) and Eq.~(3), respectively (see main text).

\begin{figure}[h!]
\centering \epsfig{file=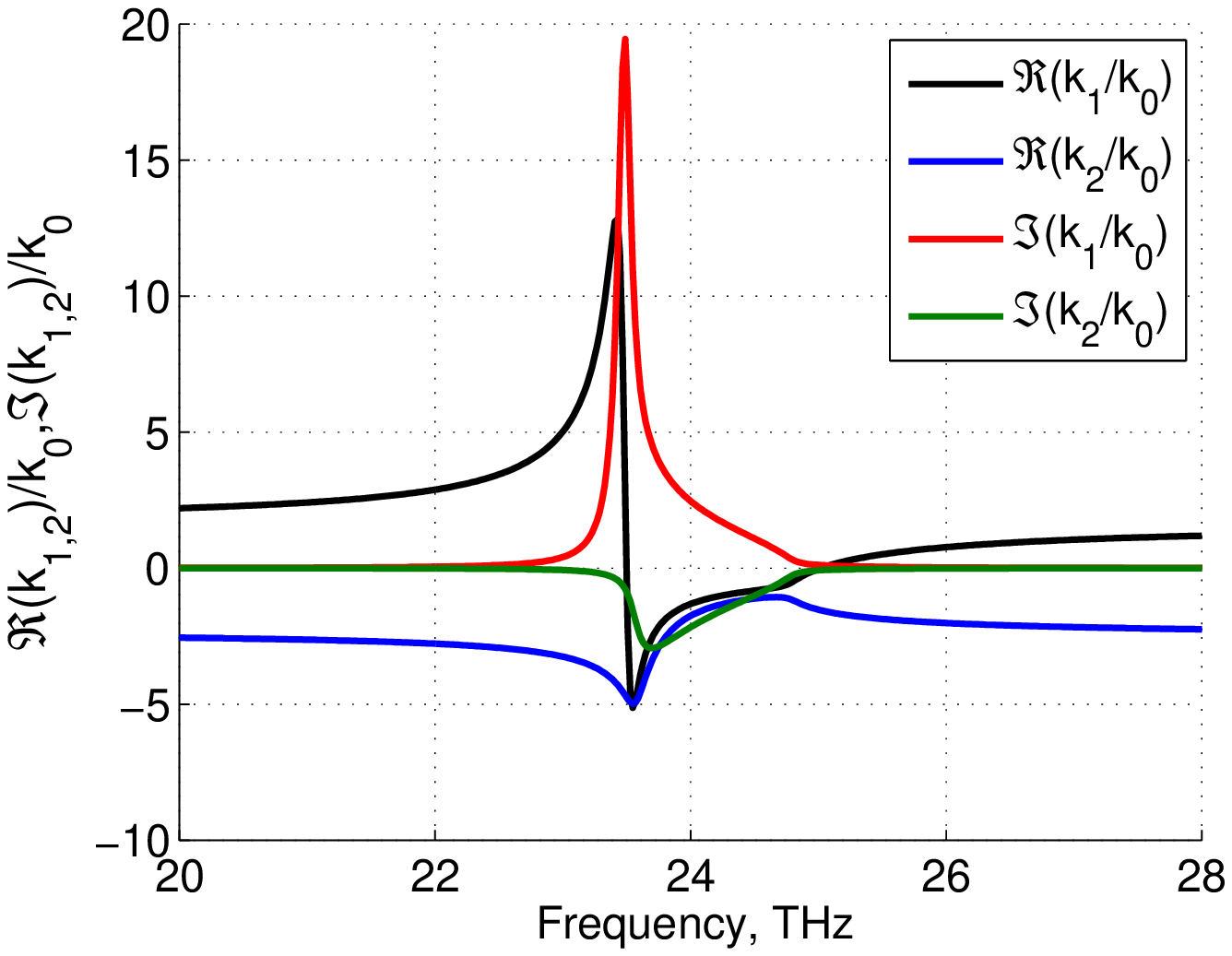, width=0.6\textwidth}
\caption{Real and imaginary parts of $k_1$ and $k_2$ versus frequency in the vicinity of $\E_{\parallel}$- resonance.}
\label{Lf}\end{figure}

\begin{figure}[h!]
\centering \epsfig{file=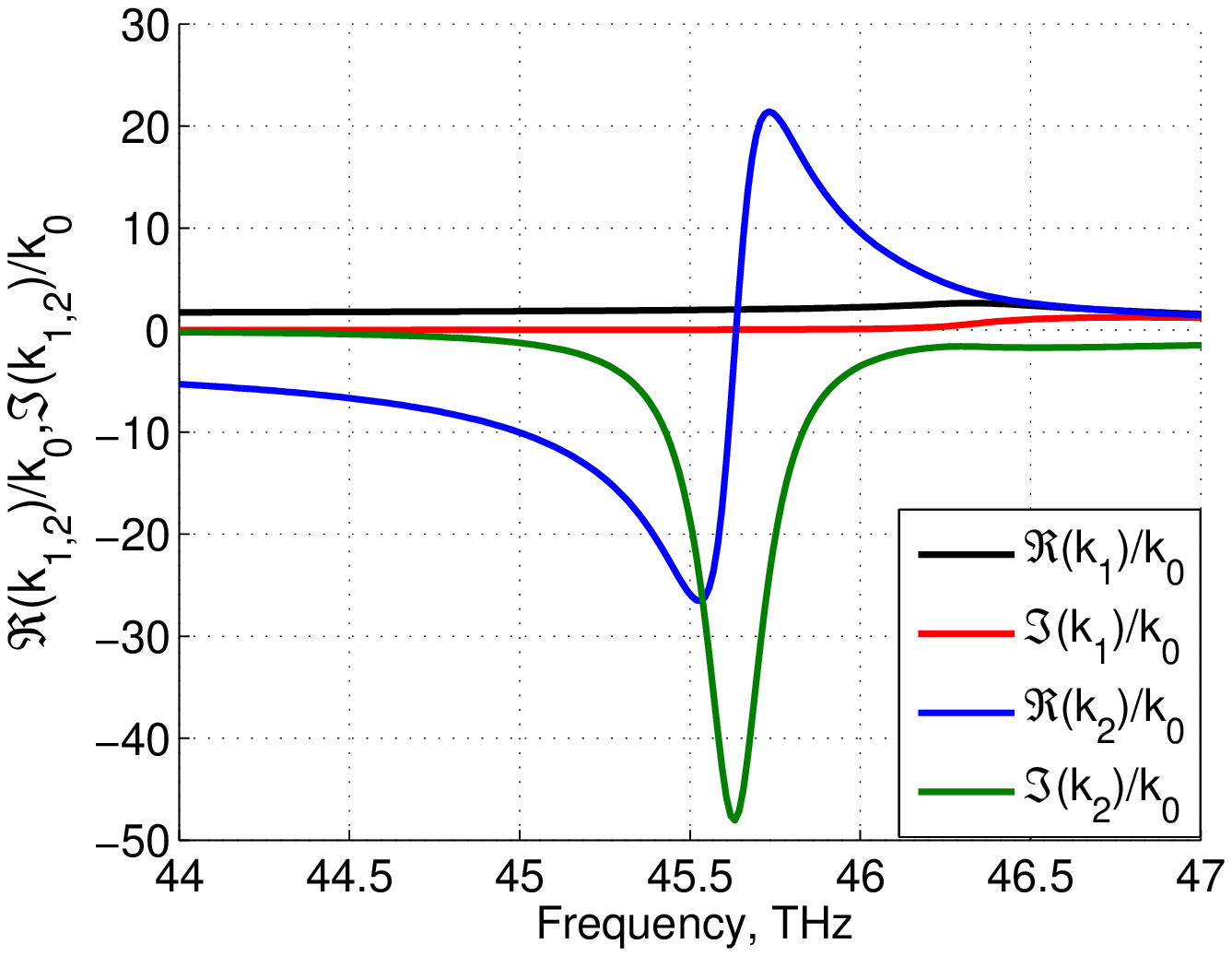, width=0.6\textwidth}
\caption{Real and imaginary parts of $k_1$ and $k_2$ versus frequency in the vicinity of $\E_{\perp}$- resonance.}
\label{Hf}\end{figure}

In order to correctly identify the waves propagating in the positive and negative directions of the $z$-axis, we have to analyze the imaginary parts of $k_z^{(1,2)}$, according to the causality principle.
Let us define the normal wave vector component for the downward wave (propagating in the positive $z$-direction) as $k_1$ and for the upward wave as $k_2$.
One then has: Im$(k_1)>0$ and Im$(k_2)<0$.
The parallel $\E_{\parallel}$ and perpendicular $\E_{\perp}$ components of the permittivity tensor of orthorhombic boron nitride, taken for illustration of the predicted effect, exhibit the Lorentzian resonances at frequencies $\approx 22.8$\,THz and $\approx 41.2$\,THz, respectively.
It is expected, that $k_1$ and $k_2$ exhibit a resonant behavior nearby these frequencies.
Fig.~\ref{Lf} illustrates the frequency dependencies of the real and imaginary parts of the normal components $k_1,\,k_2$ of wave vectors.
One can see that Re($k_1$) changes the sign in the vicinity of the resonance, so the downward propagating wave
becomes the forward wave at low frequencies and the backward wave in the frequency range from $\approx 23.6$\,THz to $\approx 25$\,THz. The upward wave remains the forward one within the considered range.
Fig.~\ref{Hf} shows similar dependencies on the frequency range around the $\E_{\perp}$ resonance.
In opposite to the previous case, here the upward wave changes the sign of dispersion.

\section{Expressions for $D_1,\,D_2$, and $D_3$}

The expressions for the $D_1,\,D_2$, and $D_3$ coefficients are found to be

\e\l{U2}\begin{array}{l}
D_1=\frac{1}{4}\left\{\frac{i|Z-Z_0|^2}{(k_{z1}-k_{z1}^*)}(1-e^{i(k_{z1}-k_{z1}^*)h})
+\frac{i|Z+Z_0|^2}{(k_{z2}-k_{z2}^*)}(1-e^{i(k_{z2}-k_{z2}^*)h})\right.-\\
\left.-\left[\frac{i(Z-Z_0)(Z^*+Z_0^*)}{(k_{z1}-k_{z2}^*)}(1-e^{i(k_{z1}-k_{z2}^*)h})+c.c.\right]
\right\}.
\end{array}
\f
\e\l{V2}\begin{array}{l}
D_2=\frac{|a|^2}{4}\left\{\frac{i|Z_0/Z-1|^2}{(k_{z1}-k_{z1}^*)}(1-e^{i(k_{z1}-k_{z1}^*)h})
+\frac{i|Z_0/Z+1|^2}{(k_{z2}-k_{z2}^*)}(1-e^{i(k_{z2}-k_{z2}^*)h})+\right.\\
\left.-\left[[\frac{i(Z_0/Z-1)(Z_0^*/Z^*+1)}{(k_{z1}-k_{z2}^*)}(1-e^{i(k_{z1}-k_{z2}^*)h})+c.c.\right]
\right\}+\\
+\frac{|b|^2}{4}\left\{\frac{i|Z_0-Z|^2}{(k_{z1}-k_{z1}^*)}(1-e^{i(k_{z1}-k_{z1}^*)h})+
\frac{i|Z_0+Z|^2}{(k_{z2}-k_{z2}^*)}(1-e^{i(k_{z2}-k_{z2}^*)h})+\right.\\
\left.+\left[\frac{i(Z_0-Z)(Z_0^*+Z^*)}{(k_{z1}-k_{z2}^*)}(1-e^{i(k_{z1}-k_{z2}^*)h})+c.c.\right]
\right\}+\\
+\left\{\frac{ab^*}{4Z}\left[\frac{i|Z_0-Z|^2}{k_{z1}-k_{z1}^*}(1-e^{i(k_{z1}-k_{z1}^*)h})
-\frac{i|Z_0+Z|^2}{k_{z2}-k_{z2}^*}(1-e^{i(k_{z2}-k_{z2}^*)h})+\right.\right.\\
\left.\left.+\frac{i(Z_0-Z)(Z_0^*+Z^*)}{k_{z1}-k_{z2}^*}(1-e^{i(k_{z1}-k_{z2}^*)h})
-\frac{i(Z_0+Z)(Z_0^*-Z^*)}{k_{z2}-k_{z1}^*}(1-e^{i(k_{z2}-k_{z1}^*)h})\right]+c.c.\right\}.
\end{array}
\f
\e\l{UV}\begin{array}{l}
D_3=\frac{1}{4}\left[\frac{iC_1(Z-Z_0)}{k_{z1}-k_{z1}^*}(1-e^{i(k_{z1}-k_{z1}^*)h})+
\frac{iC_2(Z+Z_0)}{k_{z2}-k_{z2}^*}(1-e^{i(k_{z2}-k_{z2}^*)h})-\right.\\
\left.-\frac{iC_2(Z-Z_0)}{k_{z1}-k_{z2}^*}(1-e^{i(k_{z1}-k_{z2}^*)h})-
\frac{iC_1(Z+Z_0)}{k_{z2}-k_{z1}^*}(1-e^{i(k_{z2}-k_{z1}^*)h})\right],
\end{array}\f
where
\e\l{c1c2}\begin{array}{l}
C_1=a^*Z_0^*/Z-a^*+b^*Z_0^*-b^*Z^*,\\
C_2=a^*Z_0^*/Z+a^*-b^*Z_0^*-b^*Z^*.
\end{array}\f